\documentclass[12pt,a4paper,leqno]{article}
\usepackage{graphicx}
\usepackage{amssymb}
\usepackage{epstopdf}
\title{On the gravitational field\\A suggestion about a possible experiment}
\author{by  B. Ferretti\\Dipartimento di Fisica dell'Universit\`{a} di Bologna, Italy}

\begin{document}

\maketitle

\section{Introduction}

Until rather recent years the fundamental suggestion of Einstein about the field nature of gravitational and inertial forces had not been directly confirmed by experiment. But now after the excellent work by Damour Thibault, Taylor \cite{damour:0} observation of pulsars have provided a unique opportunity for testing the field regime of relativistic gravity.

Considering however the fundamental importance of the gravity field not only at astronomical distances, but probably also for topological questions I think it would be interesting to collect new experimental evidence on the retardation of gravity interaction on the surface of the earth (see  \cite{damour:0} \cite{einstein:1} \cite{einstein:1a} \cite{veneziano:2} \cite{ferretti:3} \cite{ferretti:4} \cite{ferretti:5}).

In making the following suggestion I have tried to treasure the teachings of the research of Dameur, Thibault and Taylor. In this work had been essential the long time of observation in constant conditions. In order to achieve the same end, I think the best is to try to produce a gravitational field depending on time in a periodical way. Strictly speaking it would be necessary then an infinite time, but between $10^8$ to $10^9$ cycles will be probably sufficient. 

The ``radiators'' then will be bodies in rotational movement with constant velocity and all with the same period of rotation. It will not be too difficult to maintain these movements sufficiently uniform for the necessary time. We have instead to foresee difficulties with the ``detectors'' of the irradiated field. Such detectors will contain ``oscillators'' sensitive for the time-varying gravitational field. Now the detector oscillators will be damped and the damping will destroy periodicity, and make impossible to reach the precision necessary for our purposes.

It will be necessary then to provide in some manner to neutralize the effects of damping.

\section{The retarded gravitational interaction between two points}

As can be immediately derived from the general relativistic equations we can describe the gravitation interaction between a point $Q$ (source field) and a point $P$ (of the detector) with a simple generalization of the Newtonian expression.

Even in case of velocity of the source small compared $c$, given the existence of retardation, we have to consider different times for the laboratory ($t_\mathrm{lab}$) and for the source ($t_\mathrm{source}$). 

Let us consider the distance $\varrho$ from source to detector

\begin{displaymath}
\varrho(Q, P, t_\mathrm{lab})
\end{displaymath}

\noindent as a function of $t_\mathrm{lab}$. We suppose that $\varrho$ is a function of $t_\mathrm{lab}$ with first and second derivatives, $\dot \varrho_{t_\mathrm{lab}}$, $\ddot \varrho_{t_\mathrm{lab}}$. In order to calculate the irradiated field we have however to consider the retardation $\tau$ which is defined by

\begin{displaymath}
v_g\tau_r = \varrho(Q, P, t_\mathrm{lab} - \tau)
\end{displaymath}

\noindent where $v_g$ is the velocity of propagation of the gravitation and $\tau$ = retardation time.

But, given the assumptions, we have

\begin{displaymath}
\varrho(Q, P, t_\mathrm{lab} - \tau) = \varrho(Q, P, t_\mathrm{lab}) - \dot\varrho_\mathrm{lab}\tau + O(\tau^2_\mathrm{lab})
\end{displaymath}

\noindent where $O(\tau^2_\mathrm{lab})$ is completely negligible.

Therefore

\begin{displaymath}
\tau = {\varrho(Q, P, t_\mathrm{lab}) \over v_g}
\end{displaymath}

For getting the retarded potential we have to evaluate

\begin{displaymath}
1 \over\varrho(Q, P, t_\mathrm{lab} -\tau)
\end{displaymath}

\noindent and we can follow the same procedure by expanding $1 \over\varrho(Q, P, t_\mathrm{lab} -\tau)$ in powers of $\tau$, and neglecting $O(\tau^2_\mathrm{lab})$ we get

\begin{displaymath}
{1 \over\varrho(Q, P, t_\mathrm{lab} -\tau)} = {1 \over\varrho(Q, P, t_\mathrm{lab})} - \tau {d\over dt}\left(1 \over\varrho(Q, P, t_\mathrm{lab})\right)
\end{displaymath}

Now eventually as a first contribution we get exactly the Newtonian potential by putting in the expression of the irradiated potential $1 \over\varrho(Q, P, t_\mathrm{lab})$ in place of $1 \over\varrho(Q, P, t_\mathrm{lab} -\tau)$, and a second contribution

\begin{displaymath}
\tau \left[-{d\over dt}\left(1 \over\varrho(Q, P, t_\mathrm{lab})\right)\right] = \tau {1\over\varrho} \dot\varrho = {\dot\varrho\over\varrho}{1\over v_g}
\end{displaymath}

It is interesting to notice for the future that the retarded potential irradiated by a point moving with a velocity small compared $c$ splits quite naturally almost exactly in two parts: the first is the Newtonian potential and the second is proportional to the lab time derivative of the first one.

This second part is of course not static and therefore not Newtonian, but it cannot yet conveniently be described like a superposition of the gravitational waves, and it is probably worth to leave it by itself.

Let us call the second part of the potential ``The time retarded Newtonian potential''. The chief purpose of the present paper is to make a few suggestions to study experimentally  ``The time retarded Newtonian potential''.

As we have pointed out we are particularly interested in Newtonian potential depending periodically on time. In this case the Fourier analysis is very useful. For instance if the Fourier expansion of the potential has merely $\cos n\omega t$ terms the time derivative has only  $\sin n\omega t$ terms and the Fourier analysis is sufficient for separating the Newtonian potential and its ``retarded part'' and to give the essentials of our own analysis.

\section{The ``diapason scheme'' of the detector}

The gravitational energy which should be detected will be imparted to masses (receptive masses) of the detector itself and will be initially under the form of kinetic energy of these masses. The damping will be produced in two ways. First by passage of this energy to the ambient (external damping), secondly by transformation, in forms of internal energy not useful for the detection (internal damping). To avoid important external damping, it appears necessary to use more than one receptive mass.

We suggest two equal masses $M_D$ and $M'_D$, $M_D = M'_D$ for instance two steel spheres $\Sigma$, $\Sigma '$ of centers $C$ and $C'$ on the same horizontal plane connected by a plane spring $S_\mathrm{pr}$. We suppose that $S_\mathrm{pr}$ has a bulge (central pivot). We suppose that $S_\mathrm{pr}$ in equilibrium lies on a vertical plane and in action is circularly bended. Such a spring uniformly bended transmits a constant torque $T_D$

\begin{displaymath}
T_D = E {ab^3\over 12} {\Delta \alpha\over \Delta x}
\end{displaymath}

\noindent where for steel $E \simeq 2 10^{12}$ dynes/cm (Young Modules) $a$ is the width of $S_\mathrm{pr}$ and $b$ its thickness. $\Delta\alpha$ is the angle of bending in radiants and $\Delta x$ the length of $S_\mathrm{pr}$ under consideration. Let us call $C_0$ the middle point of $C$ $C_1$. Moreover let us call $\pi_D$ the  vertical plane for $C_0$ and normal to $CC_1$ when the receptor system $M_D$ $S_\mathrm{pr}$ $M'_D$ is not stimulated and in equilibrium position.

The irradiators of variable gravity will be disposed symmetrically with respect $\pi_D$ in such a way that $M_D$ and $M'_D$ will be equally stimulated instant by instant. This way the movements of $M_D$ and $M'_D$ and consequently the torque $T_D$ and $T'_D$ due to $M_D$ and $M'_D$ must be specularly identical and $T_D$ and $T'_D$ will make   equilibrium on the plane $\pi_D$ like in a Diapason. There is however an essential difference. In the case of Diapason the forces acting on the two prongs are equal and opposite; in our case the forces due to the radiator have the same direction and their resultant is different from zero. By good luck we can precalculate exactly this resultant (see section II) and we must compensate it for instance with an electrostatic force applied on an area of approximately $1 \mathrm{cm}^2$ at the central part of $S_\mathrm{pr}$ without perturbing the torque. What we suggest to is to measure ``torques'' and not ``forces'' in order to avoid radically the dispersion of energy due to the vincular forces.

If the detector is operated correctly the central pivot must remain fully undisturbed and therefore totally at rest. No energy should pass from the detector to the surroundings through this way. If the receptor system is closed in a good vacuum not important external damping should be produced. More important will be internal damping and namely the damping due to plastic deformation and ionic plasma excitation in $S_\mathrm{pr}$. But the drastic reduction of that reason of damping cannot be reached merely by design. For the moment we must limit ourselves to precalculate which results can be achieved if a certain number of oscillations with negligible damping will be possible.

At this point it might be useful to discuss a numerical example.

Suppose now that the moving part of the radiator is essentially consisting of two identical steel spheres $\Sigma_R$ $\Sigma_R'$ with centers $C_R$ $C_R'$ moving in diametral opposite position on a circle $\Gamma_R$. Suppose $C_{\Gamma_R}$ is the centre of $\Gamma_R$ and $x_0 = C_{\Gamma_R} - C_0$ (distance of $C_{\Gamma_R}$ from $C_0$) $C_0$ being as said the middle point of $CC_1$.

For parity reasons the required symmetry proprieties of the irradiator will be assured if all the moving parts of the irradiators but $\Sigma_R$ and $\Sigma_{R'}$ are circularly symmetrical with respect a straight line vertical for $C_{\Gamma_R}$.

Notice that $x_0 -R$ is the minimum distance of $C_R$ (or $C_R'$) from $C_0$. We assume

\begin{displaymath}
x_0 -R \gg L
\end{displaymath}

\noindent so that $x_0 -R$ will be also the minimum distance of $C_R$ from $C$ and so on.

As we said if our detector is operated correctly it is equivalent to two simple oscillators moving in a specular way. Let us now concentrate on one single oscillator by itself.

At this point we can perhaps give a numerical example.

The mass $M'_D = M_D = 102 \,\,\mathrm{kg}$ (the density of iron is $8 \,\,\mathrm{gr}/\mathrm{cm}^3$). The radius $R = 14.5 \,\,\mathrm{cm}$ $x_0 = 10 \mathrm{cm}$. The momentum of inertia of $\Sigma '$ with respect to $\varrho_D$ is given by

\begin{displaymath}
\mathcal{M}_{\Sigma '} = M'_D\left((x_0+R)^2 + {2\over 5} R^2\right)
\end{displaymath}

Let us consider now the case in which there is not irradiated gravitational field. The equation of motion of our single detector is then

\begin{displaymath}
\mathcal{M}_{\Sigma '} \ddot\alpha + T_D\alpha = 0
\end{displaymath}

\noindent where $\alpha$ is the total angle of bending of our half-spring.

We can conveniently write also

\begin{displaymath}
\ddot\alpha = - \omega_D^2\alpha
\end{displaymath}

\noindent where

\begin{displaymath}
\omega_D = \sqrt{T_D\over\mathcal{M}_{\Sigma '} }
\end{displaymath}

\noindent is the proper frequency of oscillation of the halfdetector.

Now if we are in presence of a irradiated gravitational field of frequency $\omega$ and potential $\Phi$ such that the halfdetector is perturbed by a force

\begin{displaymath}
M_D{\partial\Phi\over\partial x}\cos\omega t
\end{displaymath}

\noindent where $x$ is normal to $S_{pr}$ its equation of motion becomes

\begin{displaymath}
\ddot\alpha + \omega^2_D\alpha = M_D{\partial\Phi\over\partial x}\cos\omega t
\end{displaymath}

We solve this equation with the usual procedure writing

\begin{displaymath}
\alpha(t) = \xi(t)\sin\omega_Dt + \eta(t)\cos\omega_D t
\end{displaymath}

\noindent and imposing

\begin{displaymath}
\dot\xi(t)\sin\omega_Dt + \dot\eta(t)\cos\omega_Dt = 0
\end{displaymath}

\noindent eventually we get

\begin{eqnarray*}
\dot\xi &=& \cos\omega_Dt{M_D\over\omega_D}{\partial\Phi\over\partial x}\cos\omega t\\
\dot\eta &=& -\sin\omega_Dt{M_D\over\omega_D}{\partial\Phi\over\partial x}\cos\omega t
\end{eqnarray*}

\noindent $\eta$ gives negligible contribution and

\begin{displaymath}
\xi = {1\over 2}{M_D\over\omega_D}{\partial\Phi\over\partial x}t \,\,\, (\mathrm{if}\,\, t < {1\over \omega -\omega_D}) + \mathrm{negligible\,\,contributions},
\end{displaymath}

Clearly all that is correct until the damping in negligible. This condition imposes that if the period of damping is $\tau = t_\mathrm{Damp}$ it must be $t \le\tau$.

We have now to come back to the really essential point of our suggestion.

This is that the two simple oscillators composing our detector should move in exactly specular way; this must be the necessary and sufficient condition to insure that the left torque is exactly balanced by the right torque. Therefore the movement of each oscillator must be exactly monitored to control the achievement of this condition. I suggest to use in last instance for this purpose the same apparatus which we use for the measurement.

Let us then come to this apparatus.

\begin{figure}[htbp]
\begin{center}
\includegraphics[width=12cm]{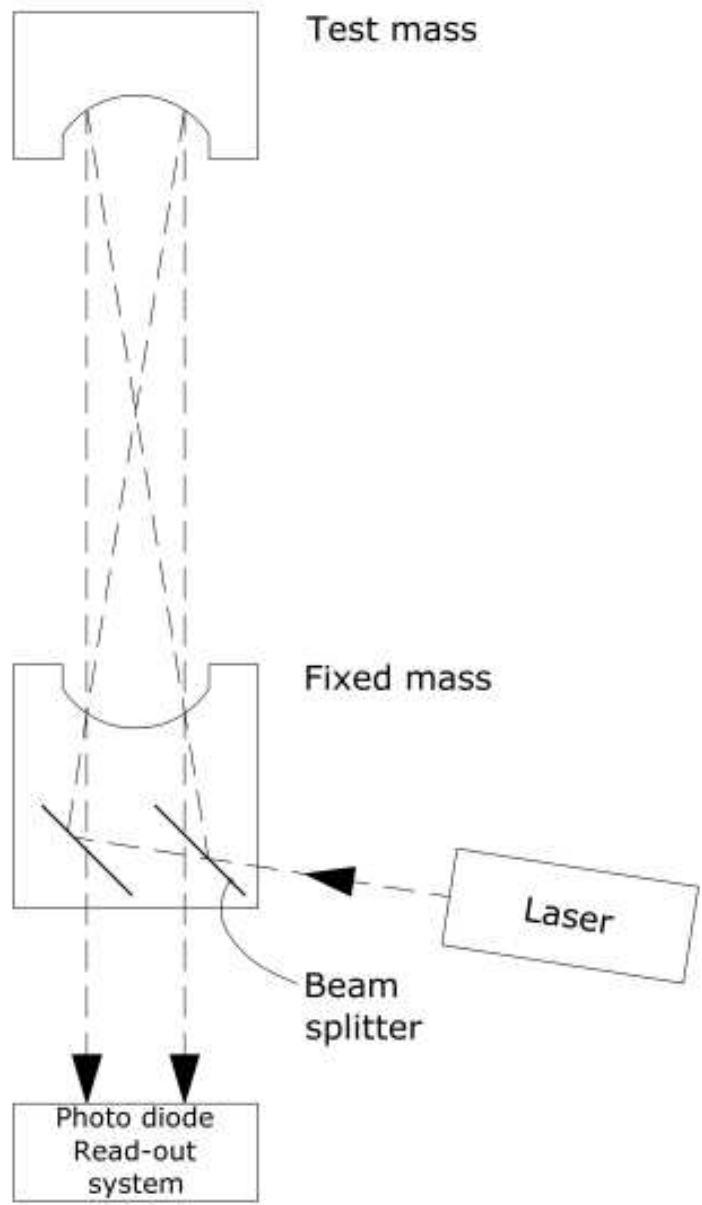}
\caption{Laser interferometer}
\label{default}
\end{center}
\end{figure}

I suggest to use a laser-interferometer similar to those used for gravitational wave detectors (see \cite{thorne:6} ). Such an interferometer allow to detect a displacement of $\sim 10^{-16}$ cm of a system of mirrors which should be connected by means of light but rigid aluminum arm to the receptive mass, for instance $M_D$, which must be monitored. In the exemplified case then a movement of $10^{-16}$ cm will be detectable. Such a sensitivity should be sufficient for our purposes. That means that the irradiators should be dimensioned and positioned with keeping these limits in mind, in making suggestions for putting in evidence experimentally the time derivative of Newtonian potentials. I think therefore that if it is decided to perform the relevant experimental work it is necessary as a first step a preliminary experimental research to see a) if its possible to obtain a ``spring'' with ``internal dissipation of energy'' sufficiently low, b) a laser interferometric device with the desired sensibility and reliability and c) a central suspension system of the detector which should support a weight of $\sim 200$ Kg s transmitting negligible small accidental torques.

Only if the preliminary problems are solved in a satisfactory way it might be reasonable to proceed to design and perform the complete experimental research.

\begin{figure}[htbp]
\begin{center}
\includegraphics[width=12cm]{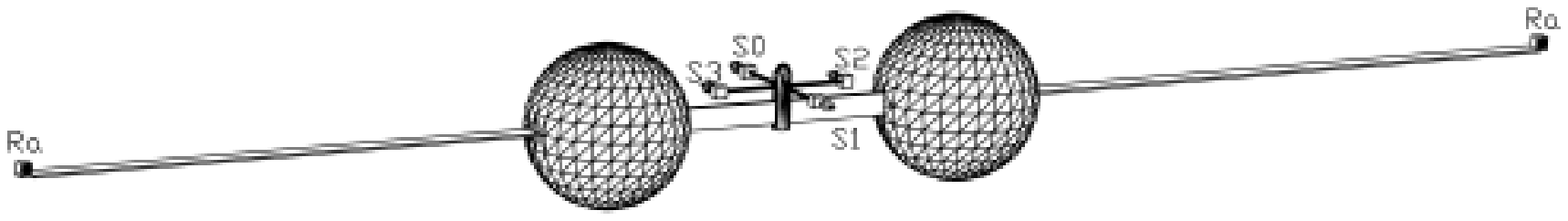}
\caption{The detector}
\label{default}
\end{center}
\end{figure}

In any case we give in fig. 3 a sketch of the central pivot of $S_{pr}$ allowing a first control of the absence of ``torques'' acting on the pivot itself.

\begin{figure}[htbp]
\begin{center}
\includegraphics[width=12cm]{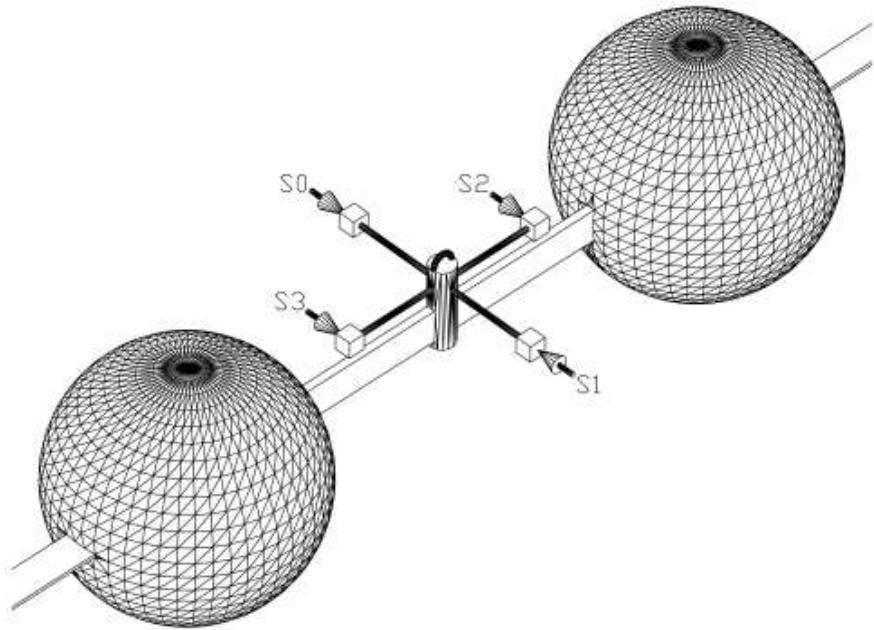}
\caption{Central pivot of the detector}
\label{default}
\end{center}
\end{figure}

\section{Appendix}

The method that we have in mind for measuring with our detector an irradiated gravitational field is a method of zero, I.e. compensating the field which must be measured with an identical but opposite field. For understanding this point is very illuminating to consider the irradiation of $n$ identical spheres of radius $r_0$ with the centers on the vertices of a regular polygon of $n$ sides of radius $R_0$ ($R_0 \gg r_0$) and center $O$. Suppose the vertices are $V_1, V_2, \ldots V_N$, then obviously the angles $V_1OV_2 = V_2OV_3 = \ldots V_{n-1}OV_1 = {2\pi\over n}$. Suppose all this system is in rotation with a constant angular velocity $\omega = 2\pi\nu$. For symmetry reasons it irradiates with a frequency $2\pi\nu n$. Suppose now we have a second identical system of spheres with centers $V'_1, V'_2, \ldots V'_N$ and suppose that the angle $V_1OV'_1 = \alpha$. If $\alpha = {1\over 2} V_1OV_2 = {1\over 2} {2\pi\over n}$ the two polygons together form a unique regular polygon $V_1V'_1V_2V'_2\ldots V'_n$ which irradiates with a frequency $2\pi\nu 2n$. That necessarily means that the ``fundamental'' radiation (of frequency $\nu$) irradiated by $V_1, V_2, \ldots V_N$ is identical, but for the sign to the ``fundamental'' radiation of $V'_1, V'_2, \ldots V'_N$. 

This gives a simple rigorous example of the method of zero we have in mind.

\textbf{Acknowledgments}

I want to thank S. Turrini and M. Pierantoni for their assistance in writing this work. Discussions with S. Bergia, E. Picasso, L.A. Radicati and R. Remiddi are also gratefully acknowledged.

Eventually I want to thank A. Passaro, Md dr, who taking care of my health, provided me with the mental and physical resources needed to carry out this work.

 \end{document}